\documentclass[twocolumn,english,aps,pra,showpacs,superscriptaddress]{revtex4-1}
\usepackage[T1]{fontenc}
\usepackage[latin9]{inputenc}
\usepackage{amsmath}
\usepackage{graphicx}
\usepackage{amssymb}
\usepackage{amsfonts}
\usepackage{amsthm}
\usepackage{multirow}

\makeatletter

\@ifundefined{textcolor}{}
{%
 \definecolor{BLACK}{gray}{0}
 \definecolor{WHITE}{gray}{1}
 \definecolor{RED}{rgb}{1,0,0}
 \definecolor{GREEN}{rgb}{0,1,0}
 \definecolor{BLUE}{rgb}{0,0,1}
 \definecolor{CYAN}{cmyk}{1,0,0,0}
 \definecolor{MAGENTA}{cmyk}{0,1,0,0}
 \definecolor{YELLOW}{cmyk}{0,0,1,0}
 }

\usepackage{txfonts}

\makeatother

\usepackage{babel}

\begin{document}

\title{Entanglement classification of $2 \times 2 \times 2 \times d $ quantum systems via the ranks of the multiple coefficient matrices}

\author{Shuhao Wang}
\affiliation{State Key Laboratory of Low-Dimensional Quantum Physics and Department of Physics, Tsinghua University, Beijing 100084,
China}
\author{Yao Lu}
\affiliation{State Key Laboratory of Low-Dimensional Quantum Physics and Department of Physics, Tsinghua University, Beijing 100084,
China}
\author{Gui-Lu Long}
\email[]{gllong@tsinghua.edu.cn}
\affiliation{State Key Laboratory of Low-Dimensional Quantum Physics and Department of Physics, Tsinghua University, Beijing 100084,
China}
\affiliation{Tsinghua National Laboratory for Information Science
and Technology, Beijing 100084, China}

\date{\today }
\begin{abstract}
The coefficient matrix is an efficient tool in entanglement classification under stochastic local operation and classical communication. In this work, we take all the ranks of the coefficient matrices into account in the method of entanglement classification, and give the entanglement classification procedure for arbitrary-dimensional multipartite pure states under stochastic local operation and classical communication. As a main application, we study the entanglement classification for quantum systems in the Hilbert space ${\cal H} = \mathbb{C}^2 \otimes \mathbb{C}^2 \otimes \mathbb{C}^2 \otimes \mathbb{C}^d$ in terms of
the ranks of the coefficient matrices.
\end{abstract}


\maketitle

\section{Introduction}
\label{intro}
Entanglement is a prime feature in quantum mechanics \cite{book}, and entangled quantum systems have given rise to novel applications in
the field of quantum information, which includes quantum teleportation \cite{bennett1993,boschi1996}, superdense coding \cite{bennett1992,zeilinger1996}, quantum computation \cite{deutsch1992,shor1994,ekert1996,grover1996},
quantum key distribution \cite{bennett1984,gisin2002}, quantum secure direct communication \cite{longliu02,beige2002,felbinger2002}, etc.
The classification of multipartite entanglement is one of the main tasks in quantum
information theory \cite{nielsen1999}. It has been shown that if two pure states $\left|\psi\right\rangle $
and $\left|\phi\right\rangle $ in ${\cal H} = \mathbb{C}^{d_1} \otimes \mathbb{C}^{d_2} \otimes \cdots \otimes \mathbb{C}^{d_n}$ are connected by stochastic local operation and classical communication (SLOCC), they can be converted into each other with the product of invertible local operators (ILOs) \begin{equation}
\left|\phi\right\rangle =F_1\otimes F_2\otimes \cdots \otimes F_n\left|\psi\right\rangle ,\end{equation}
where ${F_1},{F_2},\cdots,{F_n}$ are ILOs
in $GL(d_1,{\Bbb C})$, $GL(d_2,{\Bbb C})$, $\cdots$, $GL(d_n,{\Bbb C})$, respectively,
and they can perform the same quantum information tasks \cite{vidal2000}.
Therefore, considerable research has been conducted on multipartite
entanglement classification under SLOCC since the beginning of this century \cite{verstraete2002,lamata2006,borsten2010,miyake2003,cornelio2006,linchen2006,junlili2012,boli2012,viehmann2011,chen2006,bastin2009}.
Verstraete \emph{et al.} obtained nine SLOCC inequivalent families of quantum systems in the Hilbert space ${\cal H} = \mathbb{C}^2 \otimes \mathbb{C}^2 \otimes \mathbb{C}^2 \otimes \mathbb{C}^2$ by using group theory \cite{verstraete2002}.
Recently, Li and Li proposed the approach of entanglement classification under SLOCC for \emph{n}-qubit pure states in terms of the ranks of the coefficient matrices (RCMs) \cite{dafali2012}, and revisited the entanglement classification of four qubit quantum systems \cite{dafali2012arxiv}.

The importance of quantum states with higher dimensions is gradually recognized in recent years. It has been shown that, compared with qubits, maximally entangled
qudits violate local realism more strongly and are less affected by noise
\cite{kaszlikowski2000,jchen2001,collins2002,cheng2009,son2006,he2011}. In quantum communication, entangled qudits are
more secure against eavesdropping attacks \cite{bechmann2000,bourennane2001,cerf2002,durt2003,fpan2006},
and also offer advantages
including greater channel capacity for quantum communication \cite{fujiwara2003}
as well as more reliable quantum processing \cite{ralph2007}.
From the experimental viewpoint, the entangled qudits can be physically realized in linear photon systems \cite{moreva2006}, nitrogen-vacancy centers \cite{nv}, etc.

As a natural generalization of Li's approach \cite{wang2012}, an \emph{n}-qudit pure state in the \emph{n}-partite
Hilbert space ${\cal H} = \mathbb{C}^{d_1} \otimes \mathbb{C}^{d_2} \otimes \cdots \otimes \mathbb{C}^{d_n}$,
can be expanded as
\begin{equation}
\left|\psi\right\rangle =\sum\nolimits _{i=0}^{\prod\nolimits _{k=1}^{n}{d_{k}}-1}{{a_{i}}\left|{s_{1}}{s_{2}}\cdots{s_{n}}\right\rangle },\end{equation}
where ${a_{i}}$ are the coefficients and $\left|{s_{1}}{s_{2}}\cdots{s_{n}}\right\rangle $
are the basis states $
\left|{{s_{1}}{s_{2}}\cdots{s_{n}}}\right\rangle =\left|{s_{1}}\right\rangle \otimes\left|{s_{2}}\right\rangle \otimes\cdots\otimes\left|{s_{n}}\right\rangle $
with ${s_k} \in \{ 0,1, \cdots ,{d_k}-1\}$, $k = 1, \cdots ,n$.
The coefficient matrix ${C_{1 \cdots l,l + 1 \cdots n}}(\left| \psi  \right\rangle )$ corresponding to $\left| \psi  \right\rangle$ is constructed by arranging ${a_i}$ $(i = 0, \cdots ,\prod\nolimits_{k = 1}^n {{d_k}}  - 1)$ in ascending lexicographical order:
\begin{widetext}
\begin{equation}
{C_{1 \cdots l,l + 1 \cdots n}}(\left| \psi  \right\rangle ) = \left( {\begin{array}{*{20}{c}}
{{a_{\underbrace {0 \cdots 0}_l\underbrace {0 \cdots 0}_{n - l}}}}& \cdots &{{a_{\underbrace {0 \cdots 0}_l\underbrace {{d_{n - l}} - 1 \cdots {d_n} - 1}_{n - l}}}}\\
{{a_{\underbrace {0 \cdots {d_l} - 1}_l\underbrace {0 \cdots 0}_{n - l}}}}& \cdots &{{a_{\underbrace {0 \cdots 1}_l\underbrace {{d_{n - l}} - 1 \cdots {d_n} - 1}_{n - l}}}}\\
 \vdots & \vdots & \vdots \\
{{a_{\underbrace {{d_1} - 1 \cdots {d_l} - 1}_l\underbrace {0 \cdots 0}_{n - l}}}}& \cdots &{{a_{\underbrace {{d_1} - 1 \cdots {d_l} - 1}_l\underbrace {{d_{n - l}} - 1 \cdots {d_n} - 1}_{n - l}}}}
\end{array}} \right),
\label{cm}
\end{equation}
\end{widetext}
where $0 < l < n$.
Suppose $\sigma$ is an element in the set of all permutations of qudits $\{\sigma\}$, which gives a permutation $\{q_1,q_2, \cdots ,q_n\}$ of $\{1,2, \cdots ,n\}$. In this case, the coefficient matrix ${\cal C}_{q_1 \cdots q_l,q_{l+1} \cdots q_n}(\left| \psi  \right\rangle )$ [$C_{q_1 \cdots q_l}(\left| \psi  \right\rangle )$ for short omitting the column qudits, or even $C_{q_1 \cdots q_l}$ when the state $\left| \psi  \right\rangle $ is given] is constructed by taking the corresponding permutation $\sigma$ on Eq. (\ref{cm}).
For two \emph{n}-qudit pure states connected by SLOCC, it has been proved that the
RCMs are equal whether or not the permutation
of qudits is fulfilled on both states \cite{wang2012}.

In the previous work \cite{wang2012}, the value of $l$ is given by
\begin{equation}
l=\rm{argmax}\{{\cal P}(\emph{l})\},
\label{maxl}
\end{equation}
where\begin{equation}
{\cal P}(l) = \prod\limits_{\{ \sigma \} } {\min \{ \prod\nolimits_{k = 1}^l {{d_{q_k}},} \prod\nolimits_{k = l + 1}^n {{d_{q_k}}\} } } \end{equation}
with $d_{q_k}$ the dimension of the particle corresponding to $q_k$, $\{ \sigma \}$ is the set that contains all the permutations, which will be defined later in Eq. (\ref{permutation}).
It is obvious from Eq. (\ref{maxl}) that $l=[n/2]$ for states with each particle of the same dimension. In other words, we used only a single coefficient matrix plus all the possible permutations of its qudits in the classification procedure. By using the approach introduced in  \cite{wang2012}, we have got 22 SLOCC families in the $2 \times 2 \times 2 \times 4$ quantum system.

In this paper, we use the multiple coefficient matrices by choosing $l$ from 0 to $[n/2]$ as tools in entanglement classification, which is more powerful than a single coefficient matrix, and gave the entanglement classification procedure for arbitrary-dimensional multipartite pure states. By using the approach we have proposed, we investigate the entanglement classification of quantum systems in the Hilbert space ${\cal H} = \mathbb{C}^2 \otimes \mathbb{C}^2 \otimes \mathbb{C}^2 \otimes \mathbb{C}^d$.
The remainder of this paper is organized as follows: In Sec. \ref{sec:2}, we give the method of entanglement classification for \emph{n}-qudit pure states. In Sec. \ref{sec:3}, the entanglement classification of the $2 \times 2 \times 2 \times d$ quantum systems is obtained in terms of the RCMs. In Sec. \ref{sec:4}, we give a short summary.

\section{Method of entanglement classification for arbitrary-dimensional quantum systems}
\label{sec:2}

The biseparable criterion shown in Ref. \cite{dafali2012arxiv} can be naturally generalized to arbitrary-dimensional mutipartite pure states.

\emph{Lemma 1} (Biseparable criterion).
An $n$-qudit pure state ${\left| \psi  \right\rangle _{1 \cdots n}}$ is biseparable, i.e., it can be separated in the form of ${\left| \phi  \right\rangle _{{q_1} \cdots {q_l}}} \otimes {\left| \varphi  \right\rangle _{{q_{l + 1}} \cdots {q_n}}}$ if and only if ${\rm rank}[C_{q_1 \cdots q_l}(\left| \psi  \right\rangle )]=1$.

Bennett \emph{et al.} \cite{bennett2011} have given the definition for genuine $n$-partite
correlations, that is, a state of $n$ particles has genuine $n$-partite
correlations if it is not a product state in every bipartite cut. The deductive conclusion, namely, an $n$-qudit pure state has genuine $n$-partite
entanglement if it is not a product state in every bipartite cut, also holds.
In view of the biseparable criterion, we have the genuine entanglement criterion for $n$-qudit pure states.

\emph{Lemma 2} (Genuine entanglement criterion).
An $n$-qudit pure state is genuinely entangled if and only if the RCMs are greater than 1.

\emph{Proof}.
If the RCMs of an $n$-qudit pure state are greater than 1,
then it is not a product state in every bipartite cut.
Therefore, it has genuine $n$-partite
entanglement.\qed

With the lemmas we have obtained, it is desirable to give the following theorem.

\emph{Theorem 1.}
All the degenerate families of $n$-qudit pure states are inequivalent to
one another under SLOCC and they can be distinguished in terms of
the RCMs.

\emph{Proof.}
For an $n$-qudit pure state, several sets of entanglement can be given.
Two particles in the same set are entangled, however, they are not entangled in different sets.
We call the sets of entanglement a partition $P$. Suppose we have two different degenerate families ${\cal DF}_1$
and ${\cal DF}_2$ with partitions $P_1$ and $P_2$, respectively.
We define $M$ to be a set of particles and $M'$ is the set of the rest particles. We require $M \in P_1$ and $M  \notin P_2$.
In light of the biseparable criterion, states in ${\cal DF}_1$ can be written as ${\left| \varphi  \right\rangle _M}{\left| \phi  \right\rangle _{M'}}$
and ${\rm rank}(C_M)=1$.
However, ${\rm rank}(C_M) \ne 1$ for states in ${\cal DF}_2$ because they are not biseparable.
Therefore, degenerate families ${\cal DF}_1$ and ${\cal DF}_2$ are inequivalent under SLOCC.\qed

We have shown that the degenerate families can be fully distinguished by the ranks of $C_{q_1}, C_{{q_1}{q_2}}, \cdots , C_{{q_1}{q_2}\cdots{q_{[n/2]}}}$. Here we choose $l=1,2,\cdots,[n/2]$ because when $l>[n/2]$, it does not bring with it any new results since ${\rm rank}(C_{{q_1}{q_2}\cdots{q_l}})={\rm rank}(C_{{q_{l+1}}{q_{l+2}}\cdots{q_n}})$. By using the RCMs, the families we have obtained can be further divided into subfamilies.
In the following, we give the basic procedure of the entanglement classification for $n$-qudit pure states via the RCMs.

In order to omit the permutations that end up exchanging
rows or columns in the coefficient matrix, for $l>1$, the permutations of qudits are included in the set \cite{dafali2012,wang2012}
\begin{equation}
\{\sigma\}=\{(w_{1},u_{1})(w_{2},u_{2}) \cdots (w_{k},u_{k})\}\label{permutation}\end{equation}
where $1 \le {w_1} < {w_2} <  \cdots  < {w_k} < l+(n \mod 2)$, $l < {u_1} < {u_2} <  \cdots  < {u_k} \le n$, $k$ varies from $0$ to $l-(n \mod 2)$, and
$({w_{i}},{u_{i}})$
represents the transposition of ${w_{i}}$ and ${u_{i}}$.
The case where $k=0$ is defined as identical permutation $\sigma=I$. Each element $\sigma$ of the set $\{\sigma\}$ gives a permutation $\{q_1,q_2,\cdots,q_n\}=\sigma\{1,2,\cdots,n\}$. When $l=1$, we choose $\sigma_k=(1,k+1)$, $k=0,1,\cdots,n-1$.

Let ${_{q_1q_2\cdots q_l}{\cal F}_r^\sigma}$ represent the family of all \emph{n}-qudit
states with the coefficient matrix $C_{{q_1}{q_2}\cdots{q_{l}}}$ of rank \emph{r} under the permutation $\sigma$. Therefore, the general expression of the subfamilies is
\begin{eqnarray}
&&{_{q_1}{\cal F}_{r_1,r_2,\cdots , r_{m_1}}^{\sigma_1,\sigma_2,\cdots, \sigma_{m_1}}}\bullet {_{q_1q_2}{\cal F}_{r_1,r_2,\cdots , r_{m_2}}^{\sigma_1,\sigma_2,\cdots, \sigma_{m_2}}} \bullet \cdots \bullet {_{q_1q_2\cdots q_{[n/2]}}{\cal F}_{r_1,r_2,\cdots , r_{m_{[n/2]}}}^{\sigma_1,\sigma_2,\cdots, \sigma_{m_{[n/2]}}}}\nonumber\\
&&= {_{q_1}{\cal F}_{r_1,r_2,\cdots , r_n}^{1,2,\cdots , n}} \cap \cdots  \cap {_{q_1q_2\cdots q_{[n/2]}}{\cal F}_{r_1,r_2,\cdots , r_{m_{[n/2]}}}^{\sigma_1,\sigma_2,\cdots, \sigma_{m_{[n/2]}}}}\nonumber\\
&&={_{q_1}{\cal F}_{r_1}^{1}} \cap \cdots \cap {_{q_1q_2\cdots q_{[n/2]}}{\cal F}_{r_{m_{[n/2]}}}^{\sigma_{m_{[n/2]}}}},
\end{eqnarray}
where we have used '$\bullet$' to represent '$\cap$' and $m_l$ $(l=1,2,\cdots,n)$ are the numbers of the permutations.

To illustrate, we consider three quantum states. The $n$-partite and $d$-dimensional GHZ state has a simple expression \cite{xsliu2002}
\begin{equation}
\left| {{\rm GHZ}} \right\rangle  = \frac{1}{{\sqrt d }}\sum\limits_{i = 0}^{d - 1} {\left| {\underbrace {ii \ldots i}_n} \right\rangle }.
\end{equation}
It has been shown that all the coefficient matrices have the form of \cite{wang2012}
\begin{equation}
C=\left( {\begin{array}{*{20}{c}}
{\frac{1}{{\sqrt d }}}&0& \cdots &0&0\\
0& \ddots & \cdots &0&0\\
 \vdots & \vdots &{\frac{1}{{\sqrt d }}}& \vdots & \vdots \\
0&0& \cdots & \ddots &0\\
0&0& \cdots &0&{\frac{1}{{\sqrt d }}}
\end{array}} \right).
\end{equation}
Thus the RCMs are all $d$. Therefore the $n$-partite and $d$-dimensional GHZ state belongs to ${_{q_1}{\cal F}_{d,d,\cdots , d}^{\sigma_1,\sigma_2,\cdots, \sigma_{m_1}}}\bullet {_{q_1q_2}{\cal F}_{d,d,\cdots , d}^{\sigma_1,\sigma_2,\cdots, \sigma_{m_2}}} \bullet \cdots \bullet {_{q_1q_2\cdots q_{[n/2]}}{\cal F}_{d,d,\cdots , d}^{\sigma_1,\sigma_2,\cdots, \sigma_{m_{[n/2]}}}}$.

The four-qubit cluster state \cite{briegel2001} is defined as
\begin{eqnarray}
| \phi_4 \rangle = \frac{1}{2}(| 0000 \rangle + | 0011 \rangle + | 1100 \rangle - | 1111 \rangle).
\end{eqnarray}
It can be calculated that ${\rm rank}(C_{q_1}^{\sigma_0})={\rm rank}(C_{q_1}^{\sigma_1})$ $=$ ${\rm rank}(C_{q_1}^{\sigma_2})$ $=$ ${\rm rank}(C_{q_1}^{\sigma_3})$ $=$ $2$, ${\rm rank}(C_{q_1 q_2}^{\sigma_0'})$ $=$ $2$, and ${\rm rank}(C_{q_1 q_2}^{\sigma_1'})$ $=$ ${\rm rank}(C_{q_1 q_2}^{\sigma_2'})$ $=$ $4$, where $\sigma_0=\sigma_0'=I, \sigma_1=(1,2),\sigma_2=\sigma_1'=(1,3),\sigma_3=\sigma_2'=(1,4)$. Therefore, the four-qubit cluster state belongs to ${_{q_1}{\cal F}_{2,2,2,2}^{\sigma_0,\sigma_1,\sigma_2,\sigma_3}}\bullet{_{q_1q_2}{\cal F}_{2,4,4}^{\sigma_0',\sigma_1',\sigma_2'}}$. Noting that the four-qubit GHZ state is in ${_{q_1}{\cal F}_{2,2,2,2}^{\sigma_0,\sigma_1,\sigma_2,\sigma_3}}\bullet{_{q_1q_2}{\cal F}_{2,2,2}^{\sigma_0',\sigma_1',\sigma_2'}}$, we can conclude that four-qubit cluster and GHZ states are classified into different subfamilies via the RCMs.

The four-qubit Dicke state with two excitations can be expressed as \cite{dafali2012,stockton2003}
\begin{eqnarray}
 | 2,4 \rangle &=& \frac{1}{\sqrt{6}}(| 0011 \rangle + | 1100 \rangle + | 0110 \rangle \nonumber\\
  &+& | 1001 \rangle + | 0101 \rangle + | 1010 \rangle).
\end{eqnarray}
The direct calculation leads to ${\rm rank}(C_{q_1}^{\sigma_0})={\rm rank}(C_{q_1}^{\sigma_1})$ $=$ ${\rm rank}(C_{q_1}^{\sigma_2})$ $=$ ${\rm rank}(C_{q_1}^{\sigma_3})$ $=$ $2$, ${\rm rank}(C_{q_1 q_2}^{\sigma_0'})$ $ = $ ${\rm rank}(C_{q_1 q_2}^{\sigma_1'})$ $=$ ${\rm rank}(C_{q_1 q_2}^{\sigma_2'})$ $=$ $3$, which indicates that the four-qubit Dicke state belongs to ${_{q_1}{\cal F}_{2,2,2,2}^{\sigma_0,\sigma_1,\sigma_2,\sigma_3}}\bullet{_{q_1q_2}{\cal F}_{3,3,3}^{\sigma_0',\sigma_1',\sigma_2'}}$. Therefore, by using the RCMs, we have shown that the four-qubit cluster state, GHZ state, and Dicke state with two excitations belong to different subfamilies, namely, they cannot be transformed into each other under SLOCC.

We further illustrate our approach by studying the entanglement classification of the $2\times 2 \times 4$ quantum system via the RCMs. Since $[l/2]=1$, we only need to consider the rank of $C_{q_1}$, the expression for the subfamilies is ${_{q_1}{\cal F}_{r_1,r_2,r_3}^{I,\sigma_1,\sigma_2}}$, where $\sigma_0=I, \sigma_1=(1,2)$, and $ \sigma_2=(1,3)$.
The entanglement classification result is given in Table I, where the three particles are represented by $i,j,k$, and $d_i=d_j=2$, $d_k=4$.

\begin{table}[!htb]
\label{tab:1}
\tabcolsep 0pt
\caption{Entanglement classification of the $2\times 2\times 4$ quantum system.}
\vspace*{-12pt}
\begin{center}
\def\temptablewidth{0.48\textwidth}
{\rule{\temptablewidth}{1pt}}
\begin{tabular*}{\temptablewidth}{@{\extracolsep{\fill}}lll}
Families & Subfamilies & Representative entangled states \\
\hline
$i-j-k$ & ${_{q_1}{\cal F}_{1,1,1}^{I,\sigma_1,\sigma_2}}$ & $\left|000\right\rangle $\\
\hline
$i-jk$ & ${_{q_1}{\cal F}_{1,2,2}^{\sigma_0,\sigma_1,\sigma_2}}$ & $\left|000\right\rangle+\left|011\right\rangle $\\
\hline
$j-ik$ & ${_{q_1}{\cal F}_{2,1,2}^{\sigma_0,\sigma_1,\sigma_2}}$ & $\left|000\right\rangle+\left|101\right\rangle $\\
\hline
$k-ij$ & ${_{q_1}{\cal F}_{2,2,1}^{\sigma_0,\sigma_1,\sigma_2}}$ & $\left|000\right\rangle+\left|110\right\rangle $\\
\hline
\multirow{4}{*}{$ijk$} & ${_{q_1}{\cal F}_{2,2,2}^{\sigma_0,\sigma_1,\sigma_2}}$ & $\left|000\right\rangle +\left|111\right\rangle $\\
& & $\left|001\right\rangle +\left|010\right\rangle + \left|100\right\rangle  $\\
& ${_{q_1}{\cal F}_{2,2,3}^{\sigma_0,\sigma_1,\sigma_2}}$ & $\left|000\right\rangle +\left|011\right\rangle + \left|102\right\rangle  $\\
& ${_{q_1}{\cal F}_{2,2,4}^{\sigma_0,\sigma_1,\sigma_2}}$ & $\left|000\right\rangle +\left|011\right\rangle + \left|102\right\rangle + \left|113\right\rangle $\\
\end{tabular*}
 {\rule{\temptablewidth}{1pt}}
       \end{center}
       \end{table}

\section{Entanglement classification of the $2 \times 2 \times 2 \times d $ quantum systems}
\label{sec:3}

For $2 \times 2 \times 2 \times d$ quantum systems, according to Eq.(\ref{permutation}), when $l=1$, the permutation of qudits can be expressed as
\begin{equation}
\sigma=\{\sigma_0,\sigma_1,\sigma_2,\sigma_3\},
\end{equation}
where $\sigma_0=I, \sigma_1=(1,2),\sigma_2=(1,3)$, and $\sigma_3=(1,4)$.
In the case where $l=2$, the permutations are
\begin{equation}
\sigma'=\{\sigma_0',\sigma_1',\sigma_2'\},
\end{equation}
where $\sigma_0'=I, \sigma_1'=(1,3)$, and $\sigma_2'=(1,4)$.
The expression of the subfamilies is
\begin{equation}
{_{q_1}{\cal F}_{r_1,r_2,r_3,r_4}^{\sigma_0,\sigma_1,\sigma_2,\sigma_3}}\bullet {_{q_1q_2}{\cal F}_{r_1',r_2',r_3'}^{\sigma_0',\sigma_1',\sigma_2'}}.
\end{equation}

In order to perform the entanglement classification, let us first study the properties of the RCMs of the $2 \times 2 \times 2 \times d$ quantum systems.

\emph{Theorem 2.}
For the $2 \times 2 \times 2 \times d$ system, the maximum of ${\rm rank}(C_{q_1})$ is $\min\{d,8\}$, and the maximum of ${\rm rank}(C_{q_1 q_2})$ is 4.

\emph{Proof.}
We prove the theorem by considering the independence of the residual particles except the particle with dimension $d$. In the case where $l=1$, there exists three independent qubits, therefore, there are eight independent state vectors in the subspace. According to the definition of the coefficient matrix, the maximal RCM is $\min\{d,8\}$. In the case where $l=2$, there exists four independent state vectors in the subspace, therefore, the maximal RCM is $\min\{2d,4\}$. Since $d \ge 2$, when $l=2$, the maximal RCM is 4. \qed

\begin{figure}[]
\begin{centering}
\includegraphics[width=7.4cm]{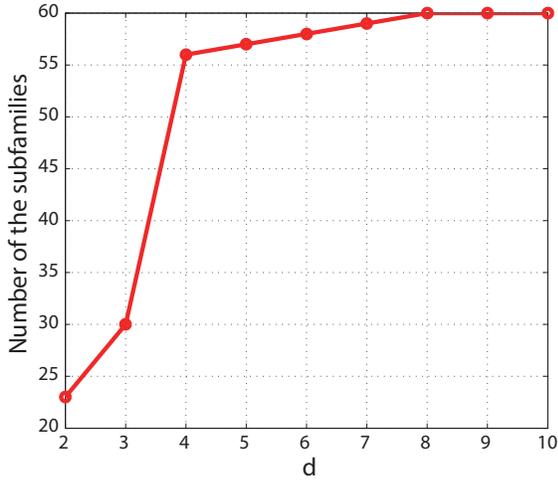}
\caption{(Color online) Dependency of the total subfamily number on $d$. When $d \ge 8$, the subfamily number keeps at 60.}
\label{fig1}
\end{centering}
\end{figure}

In order to get the maximum subfamily number, without loss of generality, we choose $d=8$. The entanglement classification of the $2 \times 2 \times 2 \times 8$ system is shown in Tables II and III, where the four particles are represented by $i,j,k,m$, and $d_i=d_j=d_k=2$, $d_m=8$.
It can be seen that there exists a total of 15 families (including 14 degenerate families), which can be further divided into 60 subfamilies.
We can summarize from Tables II and III that the total number of the subfamilies is concerned with $d$. The dependency of the total subfamily number on $d$ is illustrated in Fig. \ref{fig1}.

\begin{table*}[]
\label{tab:2}
\tabcolsep 0pt
\caption{Entanglement classification of the $2\times2\times2\times 8$ quantum system, where $\sigma_0=\sigma_0'=I, \sigma_1=(1,2),\sigma_2=\sigma_1'=(1,3)$, and $\sigma_3=\sigma_2'=(1,4)$.}
\vspace*{-18pt}
\begin{center}
\def\temptablewidth{0.75\textwidth}
{\rule{\temptablewidth}{1pt}}
\begin{tabular*}{\temptablewidth}{@{\extracolsep{\fill}}lll}
Families & Subfamilies & Representative entangled states \\
\hline
$i-j-k-m$ & ${_{q_1}{\cal F}_{1,1,1,1}^{\sigma_0,\sigma_1,\sigma_2,\sigma_3}}\bullet{_{q_1q_2}{\cal F}_{1,1,1}^{\sigma_0',\sigma_1',\sigma_2'}}$ & $\left|0000\right\rangle $\\
\hline
$i-j-km$ & ${_{q_1}{\cal F}_{1,1,2,2}^{\sigma_0,\sigma_1,\sigma_2,\sigma_3}}\bullet{_{q_1q_2}{\cal F}_{1,2,2}^{\sigma_0',\sigma_1',\sigma_2'}}$ & $\left|1010\right\rangle+\left|1001\right\rangle $\\
\hline
$i-k-jm$ & ${_{q_1}{\cal F}_{1,2,1,2}^{\sigma_0,\sigma_1,\sigma_2,\sigma_3}}\bullet{_{q_1q_2}{\cal F}_{2,2,1}^{\sigma_0',\sigma_1',\sigma_2'}}$ & $\left|1100\right\rangle+\left|1001\right\rangle $\\
\hline
$i-m-jk$ & ${_{q_1}{\cal F}_{1,2,2,1}^{\sigma_0,\sigma_1,\sigma_2,\sigma_3}}\bullet{_{q_1q_2}{\cal F}_{2,1,2}^{\sigma_0',\sigma_1',\sigma_2'}}$ & $\left|1100\right\rangle+\left|1010\right\rangle $\\
\hline
$j-k-im$ & ${_{q_1}{\cal F}_{2,1,1,2}^{\sigma_0,\sigma_1,\sigma_2,\sigma_3}}\bullet{_{q_1q_2}{\cal F}_{2,1,2}^{\sigma_0',\sigma_1',\sigma_2'}}$ & $\left|1100\right\rangle+\left|0101\right\rangle $\\
\hline
$j-m-ik$ & ${_{q_1}{\cal F}_{2,1,2,1}^{\sigma_0,\sigma_1,\sigma_2,\sigma_3}}\bullet{_{q_1q_2}{\cal F}_{2,2,1}^{\sigma_0',\sigma_1',\sigma_2'}}$ & $\left|1100\right\rangle+\left|0110\right\rangle $\\
\hline
$k-m-ij$ & ${_{q_1}{\cal F}_{2,2,1,1}^{\sigma_0,\sigma_1,\sigma_2,\sigma_3}}\bullet{_{q_1q_2}{\cal F}_{1,2,2}^{\sigma_0',\sigma_1',\sigma_2'}}$ & $\left|1010\right\rangle+\left|0110\right\rangle $\\
\hline
\multirow{4}{*}{$i-jkm$} & ${_{q_1}{\cal F}_{1,2,2,2}^{\sigma_0,\sigma_1,\sigma_2,\sigma_3}}\bullet{_{q_1q_2}{\cal F}_{2,2,2}^{\sigma_0',\sigma_1',\sigma_2'}}$ & $\left|0000\right\rangle+\left|0111\right\rangle $\\
&&$\left|0010\right\rangle+\left|0001\right\rangle+\left|0100\right\rangle $\\
& ${_{q_1}{\cal F}_{1,2,2,3}^{\sigma_0,\sigma_1,\sigma_2,\sigma_3}}\bullet{_{q_1q_2}{\cal F}_{2,3,2}^{\sigma_0',\sigma_1',\sigma_2'}}$ & $\left|0000\right\rangle+\left|0011\right\rangle+\left|0102\right\rangle $\\
& ${_{q_1}{\cal F}_{1,2,2,4}^{\sigma_0,\sigma_1,\sigma_2,\sigma_3}}\bullet{_{q_1q_2}{\cal F}_{2,4,2}^{\sigma_0',\sigma_1',\sigma_2'}}$ &$\left|0000\right\rangle+\left|0011\right\rangle+\left|0102\right\rangle+\left|0113\right\rangle $\\
\hline
\multirow{4}{*}{$j-ikm$} & ${_{q_1}{\cal F}_{2,1,2,2}^{\sigma_0,\sigma_1,\sigma_2,\sigma_3}}\bullet{_{q_1q_2}{\cal F}_{2,2,2}^{\sigma_0',\sigma_1',\sigma_2'}}$ & $\left|0000\right\rangle+\left|1011\right\rangle $\\
&&$\left|0010\right\rangle+\left|0001\right\rangle+\left|1000\right\rangle $\\
& ${_{q_1}{\cal F}_{2,1,2,3}^{\sigma_0,\sigma_1,\sigma_2,\sigma_3}}\bullet{_{q_1q_2}{\cal F}_{2,2,3}^{\sigma_0',\sigma_1',\sigma_2'}}$ & $\left|0000\right\rangle+\left|0011\right\rangle+\left|1002\right\rangle $\\
& ${_{q_1}{\cal F}_{2,1,2,4}^{\sigma_0,\sigma_1,\sigma_2,\sigma_3}}\bullet{_{q_1q_2}{\cal F}_{2,2,4}^{\sigma_0',\sigma_1',\sigma_2'}}$ &$\left|0000\right\rangle+\left|0011\right\rangle+\left|1002\right\rangle+\left|1013\right\rangle $\\
\hline
\multirow{4}{*}{$k-ijm$} & ${_{q_1}{\cal F}_{2,2,1,2}^{\sigma_0,\sigma_1,\sigma_2,\sigma_3}}\bullet{_{q_1q_2}{\cal F}_{2,2,2}^{\sigma_0',\sigma_1',\sigma_2'}}$ & $\left|0000\right\rangle+\left|1101\right\rangle $\\
&&$\left|1000\right\rangle+\left|0001\right\rangle+\left|0100\right\rangle $\\
& ${_{q_1}{\cal F}_{2,2,1,3}^{\sigma_0,\sigma_1,\sigma_2,\sigma_3}}\bullet{_{q_1q_2}{\cal F}_{3,2,2}^{\sigma_0',\sigma_1',\sigma_2'}}$ & $\left|0000\right\rangle+\left|1001\right\rangle+\left|0102\right\rangle $\\
& ${_{q_1}{\cal F}_{2,2,1,4}^{\sigma_0,\sigma_1,\sigma_2,\sigma_3}}\bullet{_{q_1q_2}{\cal F}_{4,2,2}^{\sigma_0',\sigma_1',\sigma_2'}}$ &$\left|0000\right\rangle+\left|1001\right\rangle+\left|0102\right\rangle+\left|1103\right\rangle $\\
\hline
$m-ijk$ & ${_{q_1}{\cal F}_{2,2,2,1}^{\sigma_0,\sigma_1,\sigma_2,\sigma_3}}\bullet{_{q_1q_2}{\cal F}_{2,2,2}^{\sigma_0',\sigma_1',\sigma_2'}}$ & $\left|0000\right\rangle+\left|1110\right\rangle $\\
&&$\left|0010\right\rangle+\left|1000\right\rangle+\left|0100\right\rangle $\\
\hline
$ij-km$ & ${_{q_1}{\cal F}_{2,2,2,2}^{\sigma_0,\sigma_1,\sigma_2,\sigma_3}}\bullet{_{q_1q_2}{\cal F}_{1,4,4}^{\sigma_0',\sigma_1',\sigma_2'}}$ & $\left|0000\right\rangle+\left|0011\right\rangle+\left|1100\right\rangle+\left|1111\right\rangle $\\
\hline
$ik-jm$ & ${_{q_1}{\cal F}_{2,2,2,2}^{\sigma_0,\sigma_1,\sigma_2,\sigma_3}}\bullet{_{q_1q_2}{\cal F}_{4,1,4}^{\sigma_0',\sigma_1',\sigma_2'}}$ & $\left|0000\right\rangle+\left|1001\right\rangle+\left|0110\right\rangle+\left|1111\right\rangle $\\
\hline
$im-jk$ & ${_{q_1}{\cal F}_{2,2,2,2}^{\sigma_0,\sigma_1,\sigma_2,\sigma_3}}\bullet{_{q_1q_2}{\cal F}_{4,4,1}^{\sigma_0',\sigma_1',\sigma_2'}}$ & $\left|0000\right\rangle+\left|1010\right\rangle+\left|0101\right\rangle+\left|1111\right\rangle$
\end{tabular*}
       {\rule{\temptablewidth}{1pt}}
       \end{center}
       \end{table*}

\begin{table*}[]
\label{tab:3}
\tabcolsep 0pt
\caption{(\emph{Continued.}) Entanglement classification of the $2\times2\times2\times 8$ quantum system, where $\sigma_0=\sigma_0'=I, \sigma_1=(1,2),\sigma_2=\sigma_1'=(1,3)$, and $\sigma_3=\sigma_2'=(1,4)$.}
\vspace*{-18pt}
\begin{center}
\def\temptablewidth{0.75\textwidth}
{\rule{\temptablewidth}{1pt}}
\begin{tabular*}{\temptablewidth}{@{\extracolsep{\fill}}lll}
Families & Subfamilies & Representative entangled states \\
\hline
\multirow{46}{*}{$ijkm$} & ${_{q_1}{\cal F}_{2,2,2,2}^{\sigma_0,\sigma_1,\sigma_2,\sigma_3}}\bullet{_{q_1q_2}{\cal F}_{2,2,2}^{\sigma_0',\sigma_1',\sigma_2'}}$ & $\left|0001\right\rangle +\left|0010\right\rangle +\left|0100\right\rangle +\left|1000\right\rangle $\\
& & $\left|1010\right\rangle +\left|1100\right\rangle+\left|1001\right\rangle  $\\
& ${_{q_1}{\cal F}_{2,2,2,2}^{\sigma_0,\sigma_1,\sigma_2,\sigma_3}}\bullet{_{q_1q_2}{\cal F}_{2,3,3}^{\sigma_0',\sigma_1',\sigma_2'}}$ & $\left|0000\right\rangle +\left|1100\right\rangle + \left|1112\right\rangle  $\\
& ${_{q_1}{\cal F}_{2,2,2,3}^{\sigma_0,\sigma_1,\sigma_2,\sigma_3}}\bullet{_{q_1q_2}{\cal F}_{2,3,3}^{\sigma_0',\sigma_1',\sigma_2'}}$ & $\left|0000\right\rangle +\left|1101\right\rangle + \left|1112\right\rangle  $\\
& ${_{q_1}{\cal F}_{2,2,2,2}^{\sigma_0,\sigma_1,\sigma_2,\sigma_3}}\bullet{_{q_1q_2}{\cal F}_{3,2,3}^{\sigma_0',\sigma_1',\sigma_2'}}$ & $\left|0000\right\rangle +\left|0110\right\rangle + \left|1102\right\rangle $\\
& ${_{q_1}{\cal F}_{2,2,2,3}^{\sigma_0,\sigma_1,\sigma_2,\sigma_3}}\bullet{_{q_1q_2}{\cal F}_{3,2,3}^{\sigma_0',\sigma_1',\sigma_2'}}$ & $\left|0000\right\rangle +\left|1001\right\rangle + \left|1112\right\rangle $\\
& ${_{q_1}{\cal F}_{2,2,2,2}^{\sigma_0,\sigma_1,\sigma_2,\sigma_3}}\bullet{_{q_1q_2}{\cal F}_{3,3,2}^{\sigma_0',\sigma_1',\sigma_2'}}$ & $\left|0000\right\rangle +\left|1010\right\rangle + \left|1112\right\rangle $\\
& ${_{q_1}{\cal F}_{2,2,2,3}^{\sigma_0,\sigma_1,\sigma_2,\sigma_3}}\bullet{_{q_1q_2}{\cal F}_{3,3,2}^{\sigma_0',\sigma_1',\sigma_2'}}$ & $\left|0000\right\rangle +\left|1011\right\rangle + \left|1112\right\rangle $\\
& ${_{q_1}{\cal F}_{2,2,2,2}^{\sigma_0,\sigma_1,\sigma_2,\sigma_3}}\bullet{_{q_1q_2}{\cal F}_{3,3,3}^{\sigma_0',\sigma_1',\sigma_2'}}$ & $\left|0000\right\rangle +\left|1111\right\rangle + \frac{1}{\sqrt{2}} (\left|0011\right\rangle+ \left|1100\right\rangle) $\\
&&$+ \frac{1}{\sqrt{3}} (\left|0101\right\rangle+ \left|1010\right\rangle+\left|0110\right\rangle+ \left|1001\right\rangle)$\\
& ${_{q_1}{\cal F}_{2,2,2,3}^{\sigma_0,\sigma_1,\sigma_2,\sigma_3}}\bullet{_{q_1q_2}{\cal F}_{3,3,3}^{\sigma_0',\sigma_1',\sigma_2'}}$ & $\left|0000\right\rangle +\left|1010\right\rangle +\left|1001\right\rangle+ \left|1112\right\rangle  $\\
& ${_{q_1}{\cal F}_{2,2,2,2}^{\sigma_0,\sigma_1,\sigma_2,\sigma_3}}\bullet{_{q_1q_2}{\cal F}_{2,4,4}^{\sigma_0',\sigma_1',\sigma_2'}}$ & $\left|0101\right\rangle+ \left|1010\right\rangle + \frac{1}{\sqrt{2}} (\left|0110\right\rangle+ \left|1001\right\rangle) $\\
& ${_{q_1}{\cal F}_{2,2,2,3}^{\sigma_0,\sigma_1,\sigma_2,\sigma_3}}\bullet{_{q_1q_2}{\cal F}_{2,4,4}^{\sigma_0',\sigma_1',\sigma_2'}}$ & $\left|0000\right\rangle +\left|1100\right\rangle + \left|0012\right\rangle + \left|1113\right\rangle  $\\
& ${_{q_1}{\cal F}_{2,2,2,4}^{\sigma_0,\sigma_1,\sigma_2,\sigma_3}}\bullet{_{q_1q_2}{\cal F}_{2,4,4}^{\sigma_0',\sigma_1',\sigma_2'}}$ & $\left|0000\right\rangle +\left|0011\right\rangle + \left|1102\right\rangle + \left|1113\right\rangle  $\\
& ${_{q_1}{\cal F}_{2,2,2,2}^{\sigma_0,\sigma_1,\sigma_2,\sigma_3}}\bullet{_{q_1q_2}{\cal F}_{4,2,4}^{\sigma_0',\sigma_1',\sigma_2'}}$ & $\left|0101\right\rangle+ \left|1010\right\rangle+ \frac{1}{\sqrt{2}} (\left|0110\right\rangle+ \left|1001\right\rangle) $\\
& ${_{q_1}{\cal F}_{2,2,2,3}^{\sigma_0,\sigma_1,\sigma_2,\sigma_3}}\bullet{_{q_1q_2}{\cal F}_{4,2,4}^{\sigma_0',\sigma_1',\sigma_2'}}$ & $\left|0000\right\rangle +\left|0110\right\rangle + \left|1002\right\rangle + \left|1113\right\rangle  $\\
& ${_{q_1}{\cal F}_{2,2,2,4}^{\sigma_0,\sigma_1,\sigma_2,\sigma_3}}\bullet{_{q_1q_2}{\cal F}_{4,2,4}^{\sigma_0',\sigma_1',\sigma_2'}}$ & $\left|0000\right\rangle +\left|1001\right\rangle + \left|0112\right\rangle + \left|1113\right\rangle  $\\
& ${_{q_1}{\cal F}_{2,2,2,2}^{\sigma_0,\sigma_1,\sigma_2,\sigma_3}}\bullet{_{q_1q_2}{\cal F}_{4,4,2}^{\sigma_0',\sigma_1',\sigma_2'}}$ & $\left|0110\right\rangle+ \left|1001\right\rangle+ \frac{1}{\sqrt{2}} (\left|0110\right\rangle+ \left|1001\right\rangle) $\\
& ${_{q_1}{\cal F}_{2,2,2,3}^{\sigma_0,\sigma_1,\sigma_2,\sigma_3}}\bullet{_{q_1q_2}{\cal F}_{4,4,2}^{\sigma_0',\sigma_1',\sigma_2'}}$ & $\left|0000\right\rangle +\left|1010\right\rangle + \left|0102\right\rangle + \left|1113\right\rangle  $\\
& ${_{q_1}{\cal F}_{2,2,2,4}^{\sigma_0,\sigma_1,\sigma_2,\sigma_3}}\bullet{_{q_1q_2}{\cal F}_{4,4,2}^{\sigma_0',\sigma_1',\sigma_2'}}$ & $\left|0000\right\rangle +\left|1011\right\rangle + \left|0102\right\rangle + \left|1113\right\rangle  $\\
& ${_{q_1}{\cal F}_{2,2,2,3}^{\sigma_0,\sigma_1,\sigma_2,\sigma_3}}\bullet{_{q_1q_2}{\cal F}_{3,3,4}^{\sigma_0',\sigma_1',\sigma_2'}}$ & $\left|0100\right\rangle +\left|1101\right\rangle + \left|0010\right\rangle + \left|1112\right\rangle  $\\
& ${_{q_1}{\cal F}_{2,2,2,4}^{\sigma_0,\sigma_1,\sigma_2,\sigma_3}}\bullet{_{q_1q_2}{\cal F}_{3,3,4}^{\sigma_0',\sigma_1',\sigma_2'}}$ & $\left|0000\right\rangle +\left|0111\right\rangle + \left|1102\right\rangle + \left|1113\right\rangle  $\\
& ${_{q_1}{\cal F}_{2,2,2,3}^{\sigma_0,\sigma_1,\sigma_2,\sigma_3}}\bullet{_{q_1q_2}{\cal F}_{3,4,3}^{\sigma_0',\sigma_1',\sigma_2'}}$ & $\left|0100\right\rangle +\left|0111\right\rangle + \left|1000\right\rangle + \left|1112\right\rangle  $\\
& ${_{q_1}{\cal F}_{2,2,2,4}^{\sigma_0,\sigma_1,\sigma_2,\sigma_3}}\bullet{_{q_1q_2}{\cal F}_{3,4,3}^{\sigma_0',\sigma_1',\sigma_2'}}$ & $\left|0000\right\rangle +\left|1101\right\rangle + \left|1012\right\rangle + \left|1113\right\rangle  $\\
& ${_{q_1}{\cal F}_{2,2,2,3}^{\sigma_0,\sigma_1,\sigma_2,\sigma_3}}\bullet{_{q_1q_2}{\cal F}_{4,3,3}^{\sigma_0',\sigma_1',\sigma_2'}}$ & $\left|0010\right\rangle +\left|0111\right\rangle + \left|1000\right\rangle + \left|1112\right\rangle  $\\
& ${_{q_1}{\cal F}_{2,2,2,4}^{\sigma_0,\sigma_1,\sigma_2,\sigma_3}}\bullet{_{q_1q_2}{\cal F}_{4,3,3}^{\sigma_0',\sigma_1',\sigma_2'}}$ & $\left|0000\right\rangle +\left|0111\right\rangle + \left|1012\right\rangle + \left|1113\right\rangle  $\\
& ${_{q_1}{\cal F}_{2,2,2,2}^{\sigma_0,\sigma_1,\sigma_2,\sigma_3}}\bullet{_{q_1q_2}{\cal F}_{3,4,4}^{\sigma_0',\sigma_1',\sigma_2'}}$ & $\left|0000\right\rangle +\left|1111\right\rangle+\left|0011\right\rangle+ \left|1100\right\rangle $\\
&&$+ \frac{1}{\sqrt{2}} (\left|0101\right\rangle+ \left|1010\right\rangle) + \frac{1}{\sqrt{3}} (\left|0110\right\rangle+ \left|1001\right\rangle)$ \\
& ${_{q_1}{\cal F}_{2,2,2,3}^{\sigma_0,\sigma_1,\sigma_2,\sigma_3}}\bullet{_{q_1q_2}{\cal F}_{3,4,4}^{\sigma_0',\sigma_1',\sigma_2'}}$ & $\left|0000\right\rangle +\left|0110\right\rangle + \left|1100\right\rangle+ \left|0012\right\rangle + \left|1113\right\rangle  $ \\
& ${_{q_1}{\cal F}_{2,2,2,4}^{\sigma_0,\sigma_1,\sigma_2,\sigma_3}}\bullet{_{q_1q_2}{\cal F}_{3,4,4}^{\sigma_0',\sigma_1',\sigma_2'}}$ & $\left|0000\right\rangle +\left|0110\right\rangle + \left|1101\right\rangle+ \left|0012\right\rangle + \left|1113\right\rangle  $ \\
& ${_{q_1}{\cal F}_{2,2,2,2}^{\sigma_0,\sigma_1,\sigma_2,\sigma_3}}\bullet{_{q_1q_2}{\cal F}_{4,3,4}^{\sigma_0',\sigma_1',\sigma_2'}}$ & $\left|0000\right\rangle +\left|1111\right\rangle+\left|0110\right\rangle+ \left|1001\right\rangle $\\
&&$+ \frac{1}{\sqrt{2}} (\left|0101\right\rangle+ \left|1010\right\rangle) + \frac{1}{\sqrt{3}} (\left|0011\right\rangle+ \left|1100\right\rangle)$ \\
& ${_{q_1}{\cal F}_{2,2,2,3}^{\sigma_0,\sigma_1,\sigma_2,\sigma_3}}\bullet{_{q_1q_2}{\cal F}_{4,3,4}^{\sigma_0',\sigma_1',\sigma_2'}}$ & $\left|0000\right\rangle +\left|0110\right\rangle +\left|1100\right\rangle +\left|1002\right\rangle + \left|1113\right\rangle  $\\
& ${_{q_1}{\cal F}_{2,2,2,4}^{\sigma_0,\sigma_1,\sigma_2,\sigma_3}}\bullet{_{q_1q_2}{\cal F}_{4,3,4}^{\sigma_0',\sigma_1',\sigma_2'}}$ & $\left|0000\right\rangle +\left|0110\right\rangle +\left|1101\right\rangle +\left|1002\right\rangle + \left|1113\right\rangle  $\\
& ${_{q_1}{\cal F}_{2,2,2,2}^{\sigma_0,\sigma_1,\sigma_2,\sigma_3}}\bullet{_{q_1q_2}{\cal F}_{4,4,3}^{\sigma_0',\sigma_1',\sigma_2'}}$ & $\left|0000\right\rangle +\left|1111\right\rangle+\left|0101\right\rangle+ \left|1010\right\rangle $\\
&&$+ \frac{1}{\sqrt{2}} (\left|0011\right\rangle+ \left|1100\right\rangle) + \frac{1}{\sqrt{3}} (\left|0110\right\rangle+ \left|1001\right\rangle)$ \\
& ${_{q_1}{\cal F}_{2,2,2,3}^{\sigma_0,\sigma_1,\sigma_2,\sigma_3}}\bullet{_{q_1q_2}{\cal F}_{4,4,3}^{\sigma_0',\sigma_1',\sigma_2'}}$ & $\left|0000\right\rangle +\left|1010\right\rangle + \left|1100\right\rangle+ \left|0102\right\rangle + \left|1113\right\rangle  $\\
& ${_{q_1}{\cal F}_{2,2,2,4}^{\sigma_0,\sigma_1,\sigma_2,\sigma_3}}\bullet{_{q_1q_2}{\cal F}_{4,4,3}^{\sigma_0',\sigma_1',\sigma_2'}}$ & $\left|0000\right\rangle +\left|1010\right\rangle + \left|1001\right\rangle+ \left|0102\right\rangle + \left|1113\right\rangle  $\\
& ${_{q_1}{\cal F}_{2,2,2,2}^{\sigma_0,\sigma_1,\sigma_2,\sigma_3}}\bullet{_{q_1q_2}{\cal F}_{4,4,4}^{\sigma_0',\sigma_1',\sigma_2'}}$ & $\left|0000\right\rangle +\left|1111\right\rangle+ \frac{1}{\sqrt{2}}(\left|0101\right\rangle+ \left|1010\right\rangle) $\\
&&$+ \frac{1}{\sqrt{3}} (\left|0011\right\rangle+ \left|1100\right\rangle) + \frac{1}{2} (\left|0110\right\rangle+ \left|1001\right\rangle)$\\
& ${_{q_1}{\cal F}_{2,2,2,3}^{\sigma_0,\sigma_1,\sigma_2,\sigma_3}}\bullet{_{q_1q_2}{\cal F}_{4,4,4}^{\sigma_0',\sigma_1',\sigma_2'}}$ & $\left|0000\right\rangle +\left|1111\right\rangle+\left|1112\right\rangle+ \frac{1}{\sqrt{2}}(\left|0101\right\rangle+ \left|1010\right\rangle) $\\
&&$+ \frac{1}{\sqrt{3}} (\left|0011\right\rangle+ \left|1100\right\rangle) + \frac{1}{2} (\left|0110\right\rangle+ \left|1001\right\rangle)$\\
& ${_{q_1}{\cal F}_{2,2,2,4}^{\sigma_0,\sigma_1,\sigma_2,\sigma_3}}\bullet{_{q_1q_2}{\cal F}_{4,4,4}^{\sigma_0',\sigma_1',\sigma_2'}}$ & $\left|0000\right\rangle +\left|0010\right\rangle + \left|0101\right\rangle + \left|0111\right\rangle$
 \\ & & $+ \left|1002\right\rangle+ \left|1012\right\rangle+ \left|1103\right\rangle+ \left|1113\right\rangle  $\\
  & $\cdots$ & $\cdots$ \\
  & ${_{q_1}{\cal F}_{2,2,2,8}^{\sigma_0,\sigma_1,\sigma_2,\sigma_3}}\bullet{_{q_1q_2}{\cal F}_{4,4,4}^{\sigma_0',\sigma_1',\sigma_2'}}$ & $\left|0000\right\rangle +\left|0011\right\rangle + \left|0102\right\rangle + \left|0113\right\rangle$
 \\ & & $+ \left|1004\right\rangle+ \left|1015\right\rangle+ \left|1106\right\rangle+ \left|1117\right\rangle  $
\end{tabular*}
       {\rule{\temptablewidth}{1pt}}
       \end{center}
       \end{table*}

\begin{figure*}[]
\begin{centering}
\includegraphics[width=18cm]{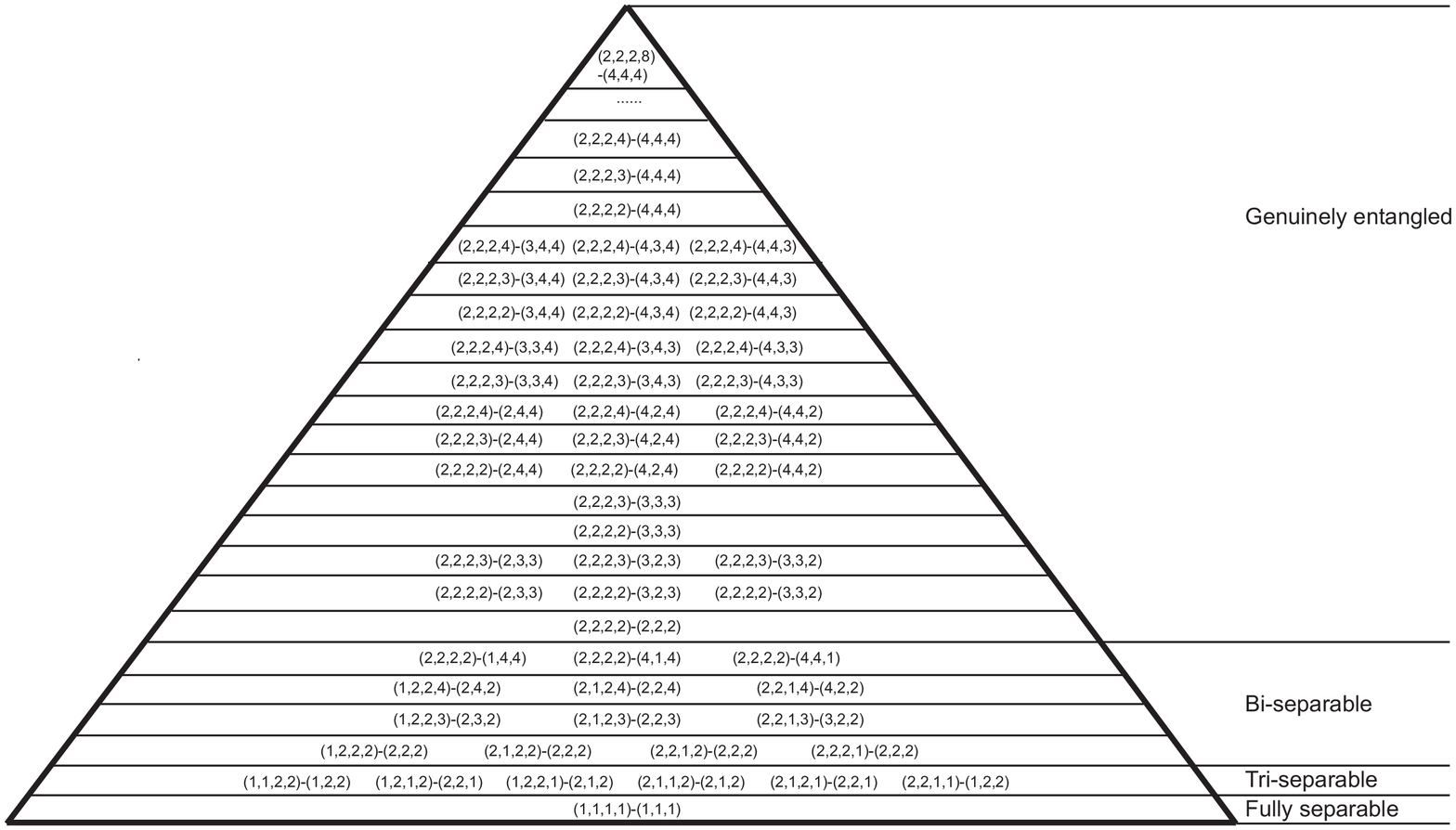}
\caption{The entanglement pyramid of the $2 \times 2 \times 2 \times 8$ quantum system, where we use $(r_1,r_2,r_3,r_4)-(r_1',r_2',r_3')$ to represent ${_{q_1}{\cal F}_{r_1,r_2,r_3,r_4}^{\sigma_0,\sigma_1,\sigma_2,\sigma_3}}\bullet {_{q_1q_2}{\cal F}_{r_1',r_2',r_3'}^{\sigma_0',\sigma_1',\sigma_2'}}$.}
\label{fig2}
\end{centering}
\end{figure*}

Noting that the RCMs are entanglement monotones \cite{wang2012,huber2013}, we illustrate the result in Tables II and III by an entanglement pyramid, which is shown in Fig. \ref{fig2}.

Figure \ref{fig2} clearly demonstrates the entanglement structure of the $2 \times 2 \times 2 \times d $ quantum systems. In general, the entanglement of the four-partite quantum systems can be classified into four types, i.e., fully separable, triseparable, biseparable, and genuinely entangled. Our physical intuition that the degree entanglement is negatively correlated to the separability of the state is consistent with the results we have obtained. Moreover, the RCMs provide us with detailed information of the entanglement of the $2 \times 2 \times 2 \times d $ quantum systems.

\section{Conclusion}
\label{sec:4}

We generalized the biseparable and genuine entanglement criteria for qubits in \cite{dafali2012arxiv} to higher dimensions (qudits). We have proved that it is possible to distinguish all the degenerate families for $n$-qudit pure states using the RCMs.
The method of entanglement classification for $n$-qudit states was given, and the entanglement classification of the $2\times 2\times 2\times d$ quantum systems was investigated. We have found at most 60 subfamilies in the $2 \times 2 \times 2 \times d$ quantum systems. In the mean time, the entanglement structure of the $2 \times 2 \times 2 \times d$ quantum systems was derived in terms of the RCMs.

It can be seen that the present approach reveals far more detailed classification results, which is far beyond the capability of the method introduced in Ref. \cite{wang2012}. If we concentrate on the case of $d = 4$, the advantage of the present approach over \cite{wang2012} is significant,  as can be seen from the entanglement pyramids in \cite{wang2012} and that in Fig. \ref{fig2}, namely, Ref. \cite{wang2012} gave only ten layers and 22 subfamilies while this work gave 22 layers and 56 subfamilies.

The reason can be explained by the entanglement of different partitions of a quantum system. In Ref. \cite{wang2012}, we only considered the partitions corresponding to a fixed $l$. However, by choosing $l=1$ to $[n/2]$, the present approach has revealed the entanglement information of all the different partitions. For instance, in the case of the $2 \times 2 \times 2 \times 4$ quantum system, Eq. (\ref{maxl}) gives $l=2$, while we choose $l=1-2$ in the the present approach. Therefore, the present approach reveals more entanglement information and gives more detailed classification results.

The procedure we have introduced can be used to study the entanglement classification of arbitrary-dimensional multipartite pure states under SLOCC. Huber and Vicente \cite{huber2013} have pointed out that the RCMs are important indicators of entanglement structure. After choosing $l = 1$ to $[n/2]$ and performing all the permutations of qudits, our approach actually gives all the independent RCMs of arbitrary-dimensional multipartite pure states. By considering the monotonicity of the RCMs, one can derive the entanglement structure via our approach.
We expect our work may find further
theoretical and experimental applications.

$ $

\section*{ACKNOWLEDGMENTS}

We thank Dafa Li for helpful discussion. This work was supported by the National Natural Science Foundation of China (Grants No. 11175094 and No. 91221205), the National Basic Research Program of China (Grants No. 2009CB929402 and No. 2011CB9216002).

\end{document}